# On Questions of Predictability and Control of an Intelligent System Using Probabilistic State-Transitions

Jayanth R Taranath

University of Texas at Austin, Department of Electrical and Computer Engineering

2501 Speedway, Austin, TX 78712

Contact: jayanth.r.t@utexas.edu

**Abstract.**

One of the central aims of neuroscience is to reliably predict the behavioral response of an organism using its neural activity. If possible, this implies we can causally manipulate the neural response and design brain-computer-interface systems to alter behavior, and vice-versa. Hence, predictions play an important role in both fundamental neuroscience and its applications. Can we predict the neural and behavioral states of an organism at any given time? Can we predict behavioral states using neural states, and vice-versa, and is there a memory-component required to reliably predict such states? Are the predictions computable within a given timescale to meaningfully stimulate and make the system reach the desired states? Through a series of mathematical treatments, such conjectures and questions are discussed. Answering them might be key for future developments in understanding intelligence and designing brain-computer-interfaces.

# 1 Premise.

An organism constantly interacts with its environment (Figure 1). We assume discrete state spaces for the organism's neural and behavioral states. The environmental state-space affects both these states and will be assumed to be latent from here onwards.

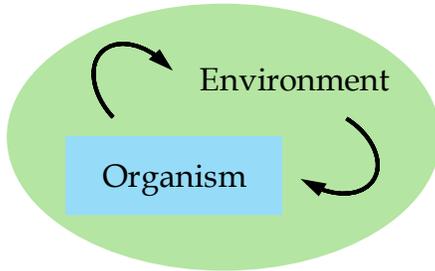

At a given time point $t$,

$N(t) \rightarrow$ Discrete Neural State
$B(t) \rightarrow$ Discrete Behavioral State
$E(t) \rightarrow$ Discrete Environmental State

$$N(t) = f(B(t), E(t))$$

$$B(t) = g(N(t), E(t))$$

$$E(t) = h(B(t)) \quad \text{E - Hidden State-Space of the Environment}$$

*Figure 1.* A schematic representation of an organism interacting with its environment and affecting each other. N(t) is a function of both the behavioral and environmental states, B(t) is a function of both the neural and environmental states, and E(t) is assumed to be a hidden state.

# 2 Probabilistic description of neural and behavioral states.

The organism's nervous system comprises of networks of neurons. These networks undergo state changes over time. We can assume that under a sufficiently high temporal resolution, the network is in a discrete state. Similarly, associated with that network state will be a behavioral state. Each neuron will be assumed to be a binary unit (0 or 1) based on the presence/absence of action-potentials (Figure 2).

We shall assume this organism has 10 neurons (hence $2^{10}$ discrete neural states) and 1000 behavioral states. Each behavioral state can be associated with multiple neural states, and vice-versa. The likelihood of observing a given neural and behavioral state pair is dependent on the hidden state (E) as well as state history (N and B before the time of observation time 't'). Hence, at any given time, there is a probability $p_t$ associated with the state-pair (Figure 2).

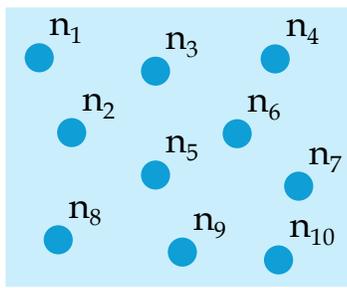

At a given time point $t$,

$N = \{n_1, n_2, n_3, n_4, n_5, n_6, n_7, n_8, n_9, n_{10}\}$; 10 neurons

Similarly,

$B_b(t) \in B(t)$ where $b \in [1,2,3,...,1000]$; 1000 possible behavioral states

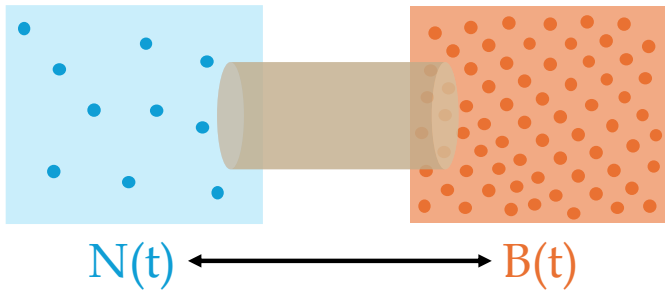

$\mathbf{P}\{N(t), B(t)\} = p_t$

Probability of having the paired states at time $t$

*Figure 2.* The joint probability of the pairs of neural and behavioral states at a given time point $t$. For demonstration purposes, we will consider 10 neurons and 1000 distinct behavioral states.

These association probabilities are computed after numerous observations of the organism's states. Once all the possible neural and behavioral states have been observed, it is important to make sure all possible associations are updated in a probability-table (Figure 3). As we do not have an analytical framework that connects N and B using E and the number of training epochs, this will be the most important step in deciding the accuracy of the decoder. Through plasticity, we also assume that the association probabilities will differ in time.

In a standard tuning-curve approach, neuronal activity is averaged over a certain number of trials and a firing-rate v/s behavior relationship is determined. These can similarly extend to population tuning-curves. But, if temporal state-transitions have an effect on the neural representations of behavior states (memory, drift, large-scale structural changes due to growth of the organism etc.) then tuning-curves will be unstable. Or rather, one cannot conceive of defining tuning-curves anymore. An advantage of the proposed approach is that such changes within the organism or the environment will not have an effect on the predicting capability as probability-tables will be updated periodically irrespective of a unit's or population representational stability.

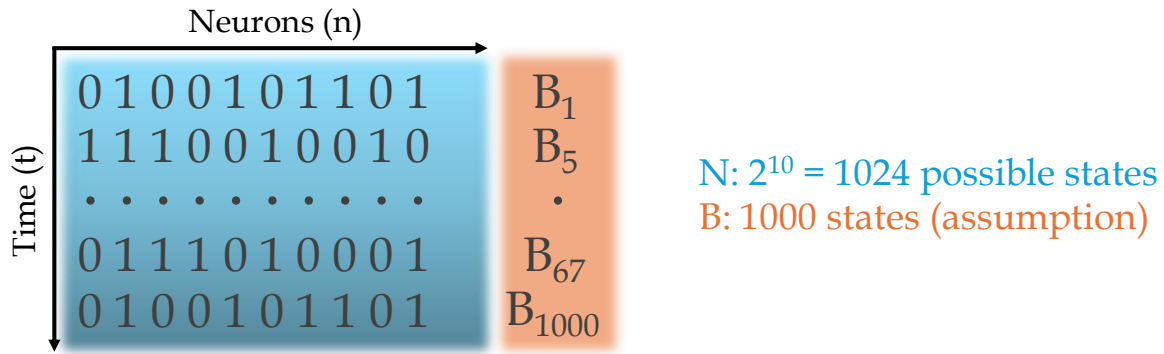

Uniqueness: $N(t) \longleftrightarrow B(t) : P\{N(t), B(t)\} = p$

Redundancy: 1) $\{N_a(t), N_b(t), N_c(t)\} \longleftrightarrow B_A(t) : \begin{array}{l} P\{N_a(t), B_A(t)\} \\ P\{N_b(t), B_A(t)\} \\ P\{N_c(t), B_A(t)\} \end{array} \Bigg] = p$

2) $N_a(t) \longleftrightarrow \{B_A(t), B_B(t), B_C(t)\} : \begin{array}{l} P\{N_a(t), B_A(t)\} \\ P\{N_a(t), B_B(t)\} \\ P\{N_a(t), B_C(t)\} \end{array} \Bigg] = p$

*Figure 3.* Uniqueness and redundancy of associations. There can be unique probabilistic states of association, or a single neuronal state can be associated with multiple behavioral states, or a single behavioral state can be represented by multiple neuronal states.

We can sort these associated states in the ascending order of the neural state indices. $(N_1, B_1): p_1$, $(N_2, [B_1, B_3]): p_2$, $(N_3, [B_3, B_5, B_7]): p_3$, until $(N_{1024}, [B_{1023}, B_{35}, \ldots B_{49}]): p_{1024}$. As mentioned above, at any given time point, one can have a table of these probabilities and their associated paired-states. If there was a full one-to-all mapping between the neural and behavioral states, the maximum bound on the size of this table would hence be $10^6$.

## 3 Transition Probabilities and State Dynamics.

At time t,

Observation: $N_t$, $B_t$, $\{N_t, B_t\}$

Probability Tables:

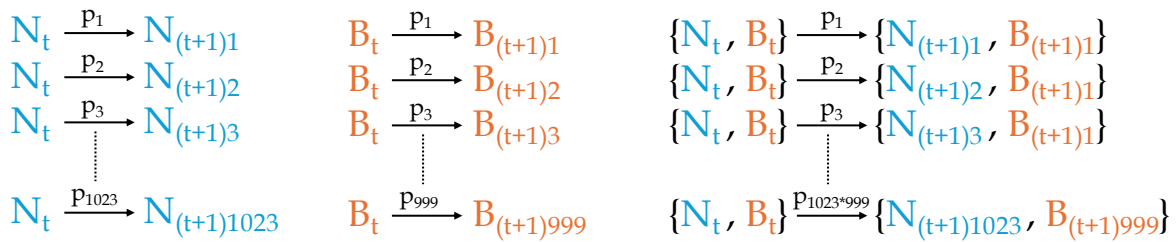

*Figure 4.* The state-transition probability tables for N, B, and {N,B} at time t updated based on data gathered during training.

How do we predict the observation of a specific neural or behavioral state (or both) at time *t + 1* given the state at *t*? If we have observed enough state-transitions of the organism, then we have the most updated probability table at time *t*. As with any learning algorithm, the number of training epochs decides the accuracy of the decoder. Here, the probability table itself becomes the kernel of the decoder, and the external observer has to choose the top-most probabilistic state-transitions as the decoder's output. In the example, at time *t*, we have observed $N_t$, $B_t$, and $\{N_t, B_t\}$ and have access to the most updated probability tables with 1023 possible state-transitions for single-states and 1,021,977 possible state-transitions for the paired-state (Figure 4).

**4 Neural-States.**

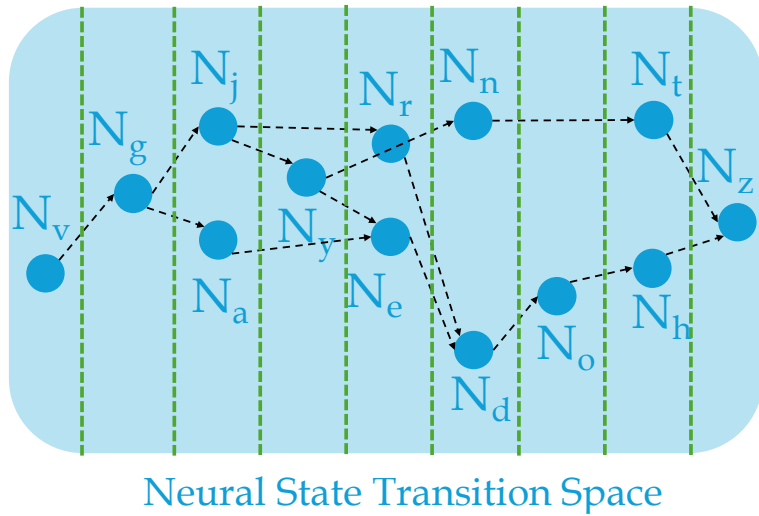

Neural State Transition Space

*Figure 5.* Visualizing neural state-transitions. Each green dashed line corresponds to the boundary of a discrete time bin.

We shall define state transition probabilities as $TP_{a,b}$ where a and b are the neural state indices (ranging from 1 to 1024). Given a state $N_a$ we can find 1023 transition probabilities corresponding to the remaining 1023 neural states. These can be called 1-step TPs (like $T_{a,b}$). Similarly, a 2-step TP can be defined as $TP_{a,b,c}$ and in general an n-step TP would be $TP_{a,b,c,\ldots,n}$ which would traverse the entire state-transition space from the start and end-times (observer-defined).

## 5 Behavioral-States.

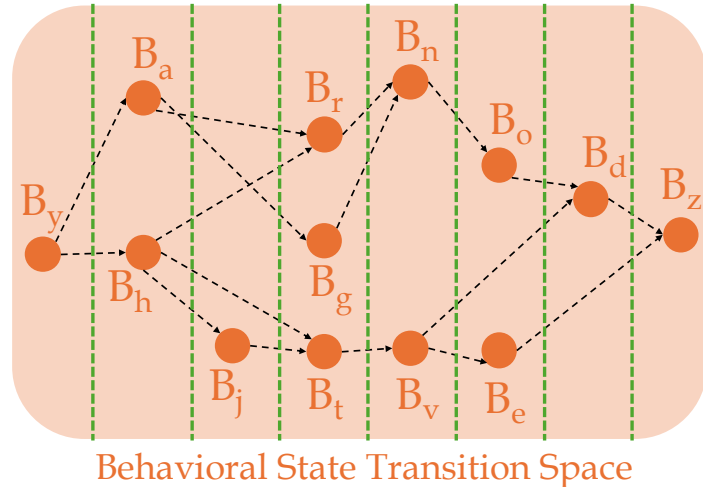

Figure 6. Visualizing behavioral state-transitions. Each green dashed line corresponds to the boundary of a discrete time bin.

The behavioral state-transitions have these crucial differences:

a) Behaviors have a more constrained set of possible transitions and thus increase the capability of predicting n-step TPs compared to a comparable neural multi-step TP (and even paired state multi-step TPs).

b) TP calculations and predictions are more practicable compared to neural and paired-state predictions, considering the current state-of-the-art image-processing and object-recognition technologies [1].

**6 Paired-States.**

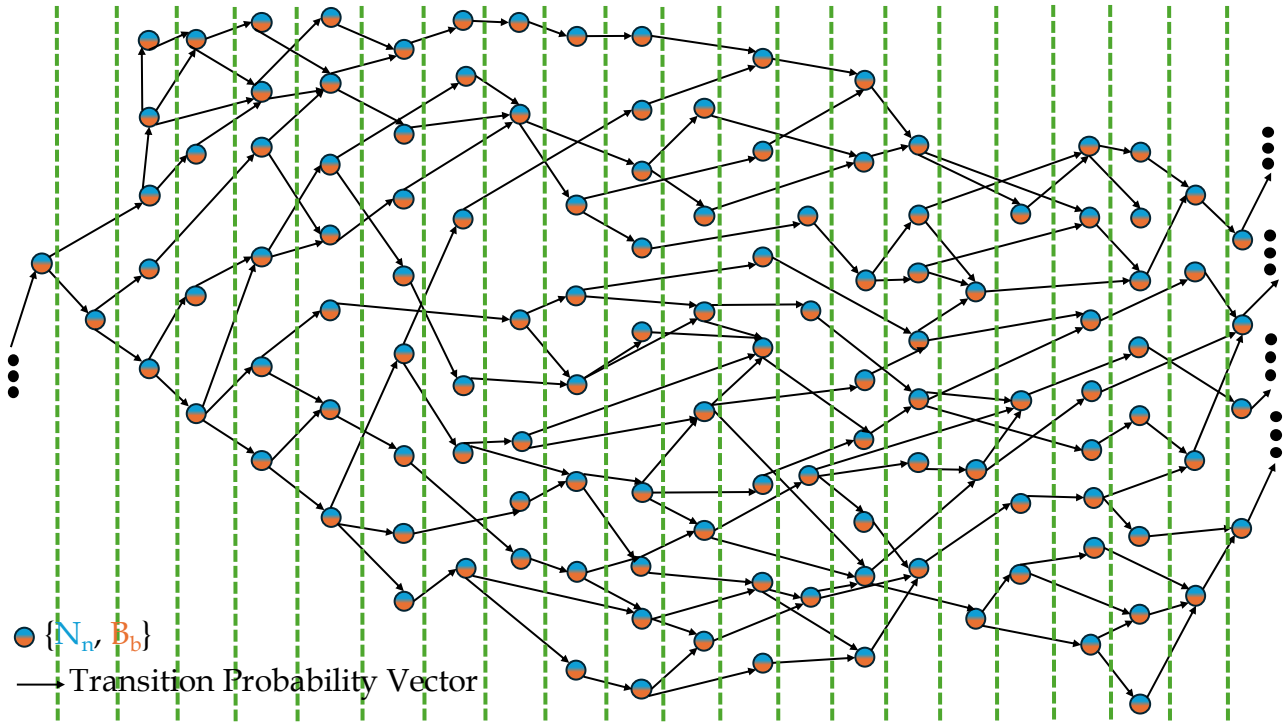

*Figure 7.* An example transition graph where each paired state is causally linked to one or multiple paired states with a specific transition probability value, defined as transition probability vectors. Each green dashed line corresponds to the boundary of a discrete time bin. Computations are performed in these individual time bins based on the evidence accumulated.

Similarly, at each time-bin, a paired-state {N, B} can be observed and the corresponding transition-probabilities can be calculated. These states can be thought of as one-half of the individual neural or behavioral states. But, due to degeneracy where multiple behavioral states can be associated with a single neural state and vice-versa, the probability-table will likely consist of a lot more state transitions than their individual neural or behavioral counterparts.

**7 An ideal state-transition prediction and perturbation machine.**

We define an ideal state-transition prediction machine as one that can, within an allowed temporal delay, *observe* in real-time the neural and behavioral states of the organism, *compute* the transition-probabilities and perturbation parameters, and *stimulate (if required)*. Now, we shall mathematically define each of these aspects and discuss the implications.

Let $t_a$ be the time-bin at which states $N_a$ and $B_a$ occur, and $t_b$ be the time-bin at which states $N_b$ and $B_b$ occur (these time-bins depend on the sampling-rate of the machine),

$\Delta m_a$ be the time required by the machine to detect the state since its occurrence,

$\Delta p_a$ be the time required by the machine to calculate and update the probability-table since $\Delta m_a$,

**Δtp$_a$** be the time required by the machine to calculate the state-transition graph until t$_S$ (S is a user-controlled parameter of the machine) since Δp$_a$,

**Δd$_a$** be the time required by the machine to compute a decision since Δtp$_a$ – to stimulate or not stimulate before the next time-bin,

**Δs$_a$** be the time required by the machine to send a stimulation signal to change and observe the changed state/s, since Δd$_a$,

Then the ideal machine should respect the following inequality:

$$t_b - t_a > t_a + \Delta m_a + \Delta p_a + \Delta tp_a + \Delta d_a + \Delta s_a$$

As with any decoder, this ideal machine will be trained on numerous state-transitions and validated for its prediction accuracy. But consider this scenario; at $t_b$, $N_b$ (let us only consider neural states) was going to be the actual state, but the state predicted by the machine was $N_c$ and a stimulation was performed to reach $N_x$ (desired state) within S. This might potentially have two effects:

a) The pre-emptive stimulation could increase the time-steps required to reach $N_x$ because of the faulty prediction.

b) If there are eventual *attractor basins* that can have adverse effects on the organism before reaching $N_x$, then the pre-emptive decision to stimulate can be damaging to the organism. Although such things can be learned through trail-and-error methods, reaching certain states even momentarily can be detrimental in the long run.

**8 Effects of noise.**

Noise in this approach can be defined as the fluctuations of neural states that contaminate the neural-behavioral state probabilities, and hence the associated state-transitions. This can be inherent to the organism (or state-dependent). Another kind that might be fundamental to the organism-machine interface is the channel noise. This would correspond to the error-rate of reading the ground-truth neural and behavioral states. The latter can have fundamental bounds explained by the physical characteristics of the interface, while the former is an inherent attribute of the organism.

**9 T-metric.**

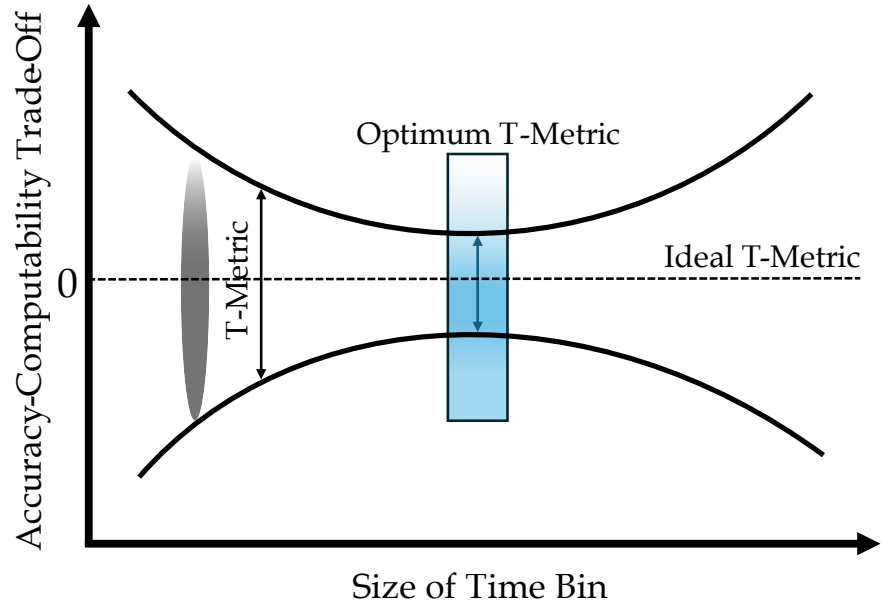

*Figure 8.* A representative graph showing the relationship between T-metric and the size of time bins (sampling rate).

We shall define T-metric as the ratio of computability to accuracy of the decoder for a given sampling rate of the machine.

$$\text{T-Metric} = \frac{\text{Computability}}{\text{Accuracy}} = [0,1]$$

Computability is defined as the extent of using the least number of decoder clocks to perform calculations within the time constraints defined by $\Delta m_a, \Delta p_a, \Delta t p_a, \Delta d_a,$ and $\Delta s_a$. Accuracy is based on the predicted immediate next-states and the state-transition graph. Important factors such as latency requirement (maximum allowed time delay between detection and decision) and branching-factor in the state-transition graph will be affected by the choice of T-metric. A two-sided computability-accuracy trade-off can be calculated for a given T-metric and an ideal machine will have a T-metric of 0. But practically, a machine should be designed with its T-metric in the optimum range shown in the blue-shaded box (Figure 8).

**10 Open Questions.**

**Q1.** What is the maximum step-number 'S' up to which $N_n$, $B_b$, or $\{N_n, B_b\}$ can be reliably predicted?

> For example, if $B_1$ to $B_{17}$ transition can only occur through the states $B_2$ and $B_5$, then $TP_{1,17} = TP_{1,2} * TP_{2,5} * TP_{5,17}$

> This forms a unique state-transition path. A reliable prediction is one that has a state-transition path with the highest TP value. After a certain S, TPs become low and equal with multiple state-transitions and a divergence is observed. Then we define the corresponding state at t = S to be the most reliably predictable state/s.

**Q2.** Does the system's n-step TP depend on the initial neural state N (t = 0)? Here, t = 0 means the very first neural state i.e., the birth state of the organism. If the number of neural states increases with increasing complexity of the organism, does having more training epochs help in the performance of the decoder? Or does the point where the training starts have a larger influence on the performance of the decoder?

**Q3.** Where should the perturbation take place in time ($t_p$) and what should be the output state of the perturbation to reliably reach a state $N_p$, $B_p$, or {$N_p$, $B_p$} within a maximum S?

**Q4.** Will information of the hidden states (E) increase S? If so, how?

**Q5.** Is it possible to compute the n-step TPs and perturbation parameters required along the path to reach the desired state within the next time bin *t+1*? What is the maximum perturbation-buffer time available outside which fresh computations of the probability-table, TPs, and perturbation parameters need to take place for successful transitions towards the desired state?

**Q6.** What is the optimum T-metric for an organism? How does it change with time (growth of the organism) and the latent environmental state-space (E)?

**Q7.** In terms of experimental validation, can C. elegans be used as the model organism to derive the temporal constraints as described for an ideal state-transition prediction machine? Recent efforts in imaging the entire neural network of the worm during active behavior and its massive parallelization could open avenues in this regard [2] [3] [4].

**Q8.** In all the previous illustrations only a single step transition-probability was considered for a given state-space. Could we increase the accuracy predictions by using multi-step transition-probabilities and storing them in the probability-tables [5]? If so, what should be the number of steps? Is there a closed-form expression for this optimum number considering it is a dynamically varying threshold? Further, will this measure influence the computability and temporal constraint parameters in a non-deterministic way?

**11 Other Approaches.**

Numerous approaches have been presented to decode the neural correlates of behavior. Their brief descriptions and how they compare with the presented approach are explained below.

**Bayesian decoding.** In classical bayesian approach, during the training phase, the prior distribution is constructed based on the statistical distribution of N and B (as before, N and B correspond to neural and behavioral states, respectively). If the goal is to find p(N|B), all we need is p(B), p(B|N), and p(N) [10]. At each time point, one gets a distribution of p(N|B) which is independent of the occurance of N and B in the previous time bin/s. If one uses a hidden-markov-model based multi-step bayesian decoding, even then one can only have information of the previous time-step's observed state and its latent states. Hence, a temporal transition-probability based method can solve this problem as one will always have access to the probability-tables at each time point with the user-defined choice of step-numbers. Further, if one decides to use the bayesian formalism for a purely neural-state-space decoder, then the causality problem arises. Here is an example. Let $p(N_a|N_b)$ denote the probability of occurance of $N_a$ given $N_b$ (note that in the traditional case, numerator and denominator should constitute different measurables as they need to be accessed simultaneously). To calculate this measure, one needs $p(N_a)$ – independent probability of occurance of $N_a$, $p(N_b)$ – independent probability of occurance of $N_b$, and $p(N_b|N_a)$ – which is the probability of occurance of $N_b$ given $N_a$. Here arises the fundamental problem of calculating how likely is the occurance of a certain neural state in the next time bin/s based on the transitions that have happened previously. So, there is a causal direction in which neural states evolve (you can consider purely behavioral states, or the neural and behavioral state trasitions combined which further complicates the computations). This information is lost in the traditional bayesian formalism.

**Pattern recognition methods.** In certain situations, kernels/ patterns of neural activity are learned by a decoding system and used in real-time to predict the behavioral responses [11]. They are especially used in systems that deal with coarser neural signals such as EEG, MEG, and fMRI and are usually followed by the application of a bayesian decoder. As mentioned above, the problem of not being able to predict state-transitions in time and in various steps persists. But it should also be noted that pattern recognition forms an inevitable pre-processing step regardless of the decoding approach. For example, even in the probabilistic state-transition approach, we need to associate a certain neural or behavioral state with a specific tag ($N_x$ or $B_x$) and the tags should be considerably different between the states. This can only be achieved by pattern recognition.

**Closed-form/ Analytical models.** Can one have a closed-form model that encompasses all the possible state-transitions of a system? This might seem too fantastical, and indeed if such a model exists, then that would be the ideal condition reached by the probabilistic state-transition approach.

**Dimensionality reduction.** Similar to pattern recognition based methods, reducing the dimensions of large-scale neural or behavioral data can only help in associating a certain state with the given metric's distribution [12]. After that first step, the open question of how well such a pre-processing system can help us predict state-transitions will remain. Perhaps, certain geometric entities (manifolds) might be highly correlated with certain patterns of neural/behavioral activity and traversing specific regions of such entities might help in predicting how the system will behave in time. But that still remains an open question and as inherently the information has been compressed into lower-dimensions, such approximations might lead to errors in predicting state-transtions.

## 12 Conclusion.

I present a state-space based approach for understanding the fundamentals of predictability and control of intelligent systems. As an illustration, we considered a simple organism with 10 neurons and a 1000 behavioral states. Probability-tables and transition-probabilities are the crucial ingredients of this method. I introduced perturbations to not only help control the organism's neural states and behavior but also to sample various state-transitions that might not happen in a given environment. I further introduced temporal constraints pertaining to the computations of state-prediction which led to conceptualizing an ideal state-prediction machine.

Numerous additions can be incorporated into this approach and tested. For example, we can explore the influence of sampling rate on the machine's computational efficiency and prediction accuracies. Perhaps, an optimal sampling rate as well as a spatial resolution (instead of neurons, it can be a cluster of neurons or entire regions of the nervous system) could help us narrow down on the best designs of the machine. Artificial neural networks can be used to predict the states and their transitions [6]. We can also envision approaching the state-transition graph as a form of bayesian network [7]. A thorough investigation along this research program might help us understand intelligence better by asking new questions and also thinking about designing the next-generation brain-computer-interfaces.

The presented approach deals with an ideal state-transition prediction machine. Probability tables are its major components. This article is not concerned with how feasible such a system is in practice as especially when one considers even a small organism such as C.elegans, one is left with millions of possible states and state-transitions. But the questions posed in this article, and the fundamental constraints that have been put-forth should help us design better decoding and BCI systems. It should also help us be cautious about claiming any inferences from such systems as we can see that the chaotic behavior of perturbations can sometimes lead to unwanted states that can be detrimental and possibly irreversible.

**Acknowledgements.**

I would like to thank Xue-Xin Wei and Pramod R Taranath for their invaluable discussions, insights, and feedback.

**Appendix.**

As per the approach introduced in this article, there is no need to keep track of how the neural connectome is changing. Although synapses weaken and strengthen, and neurotransmitters can influence activations of neurons far off and non-connected from the obvious candidate neurons, we are only considering a connection-agnostic state-space and state-transitions. Is there a way to incorporate the connectome, synaptic plasticity, neurotransmitters etc. into the probabilistic state-transition framework?

As behaviors can only be discretized and are in reality continuous variables in time, an optimum temporal resolution is required for practical implementations. If the temporal resolution is too low, then there is an over-approximation of states and their transitions, and if it is too high, then unnecessary or redundant information might create difficulties in storing and updating the probability-tables. This is not a problem with neural states as they have been assumed to be inherently binary. Even with behaviors, discretization might not be difficult with simpler organisms owing to a small behavior space but becomes a crucial issue with complex organisms.

**Perturbation as a means of spanning the state-transition space.**

The natural state transitions of an organism within an environment can be sampled if measured for long enough durations. But, what if the environment is changed (the hidden-state dependency of the neural and behavioral states changes)? We can conjecture that external perturbations, especially during the training period, can be used to simulate such hidden-state dependency changes. Hence, perturbation can become a tool to sample points in the global environment space of the organism. We can have both neural and behavioral perturbations. A neural perturbation's amplitude can be defined based on the number of individual neural units (neurons, clusters, or regions) affected with a complete inversion of the binary value (0 to 1 or 1 to 0) at the time of perturbation. It is important to note that there is no way of comparing the effect of perturbation with a reference, as once a perturbation is done and the state is changed, the state-transitions that occur thereafter will be completely independent of any hypothetical transitions that might have occurred sans-perturbation. This is akin to the measurement problem found in quantum mechanics where once the wavefunction collapses there is no way of determining the effects of the alternate states and their unravelling in time [8].

**Statistical mechanics of the system.**

A potential future exploration can be on formalizing the problem in terms of the basic equations of statistical mechanics. Each neuron can be described as a 2-value activity vector. At each time point $t$, the organism's neural state can be considered probabilistic and hence a Boltzmann's equation formalism would be appropriate [9]. Naturally, collective measures such as temperature, entropy, and density of states can be defined which might be useful in designing the state-transition prediction machine. A complete mathematical treatment is left for a future article.